\documentclass[preprint,superscriptaddress]{revtex4}
\usepackage[dvips]{graphicx}
\begin{document}

\title{Silicon-based spin and charge quantum computation}
\author{Belita Koiller}
\affiliation{Instituto de F\'{\i}sica, Universidade Federal do Rio de
Janeiro, Cx. Postal 68.528, Rio de Janeiro, 21945-970, Brazil}
\author{Xuedong Hu}
\affiliation{Department of Physics, University at Buffalo, SUNY, Buffalo, NY
14260-1500}
\author{R.B.Capaz}
\affiliation{Instituto de F\'{\i}sica, Universidade Federal do Rio de
Janeiro, Cx. Postal 68.528, Rio de Janeiro, 21945-970, Brazil}
\author{A.S.Martins}
\affiliation{Instituto de F\'{\i}sica, Universidade Federal Fluminense, 
Niter\'oi, 24210-340, Brazil}

\author{S. Das Sarma}
\affiliation{Condensed Matter Theory Center, Department of Physics, University
of Maryland, College Park, MD 20742-4111}

\begin{abstract}
Silicon-based quantum-computer architectures have attracted
attention because of their promise for scalability and their potential
for synergetically utilizing the available resources associated with the
existing Si technology infrastructure.  Electronic and nuclear spins of 
shallow donors (e.g. phosphorus) in Si are ideal candidates for 
qubits in such proposals due to the relatively long spin coherence times. 
For these spin qubits, donor electron charge manipulation by external gates 
is a key ingredient for control and read-out of single-qubit operations, while 
shallow donor exchange gates are frequently invoked to perform two-qubit 
operations. More recently, charge qubits based on tunnel coupling in P$_2^+$ 
substitutional molecular ions in Si have also been proposed. 
We discuss the feasibility of the building blocks involved in shallow donor 
quantum computation in silicon, taking into account the peculiarities of
silicon electronic structure, in particular the six degenerate states at the
conduction band edge.  We show that quantum interference among these states
does not significantly affect operations involving a single donor, but leads
to fast oscillations in electron exchange coupling and on tunnel-coupling
strength when the donor pair relative position is changed on a
lattice-parameter scale.  
These studies illustrate the considerable potential as well as the tremendous 
challenges posed by donor spin and charge as candidates for qubits in
silicon.  

\end{abstract}
\maketitle

\pagebreak

\section{Introduction}

Most of the computer-based encryption algorithms presently in use to protect 
systems accessible to the public, in particular over the Internet, rely on the
fact that factoring a large number into its prime factors is so
computationally intensive that it is practically impossible.  These systems
would be vulnerable if faster factoring schemes became viable.  The
development by Shor, about a decade ago, of a quantum algorithm that can
factorize large numbers exponentially faster than the available classical
algorithms \cite{Shor} thus could make the public key encryption scheme
potentially vulnerable, and has naturally generated widespread interest in
the study of quantum computing and quantum information processing
\cite{nielsen00,Reviews}.  The exponential speedup of Shor's algorithm is due
to the intrinsic quantum parallelism in the superposition principle and the
unitary evolution of quantum mechanics.  It implies that a computer made up
of entirely quantum mechanical parts, whose evolution is governed by quantum
mechanics, would be able to carry out in reasonably short time prime
factorization of large numbers that is prohibitively time-consuming in
classical computation, thus revolutionizing cryptography and information
theory.  Since the invention of Shor's factoring algorithm, it has also been
shown that error correction can be done to a quantum system \cite{error}, so
that a practical quantum computer (QC) does not have to be forever perfect 
to be useful, as long as quantum error corrections can be carried out.  
These two key mathematical developments have led to the creation of the new
interdisciplinary field of quantum computation and quantum information.

The elementary unit of a QC is the quantum bit, or qubit, which is a two-level
quantum system ($|0\rangle$ and $|1\rangle$).  Contrary to a classical bit
which is in one of the binary states, either 0 or 1, the state of a qubit
could be any quantum-mechanical superposition state of this two-level system:
$\alpha |0\rangle + \beta |1\rangle$, where $\alpha$ and $\beta$ are complex
numbers constrained to the normalization $|\alpha|^2 + |\beta|^2 =1 $.  The
computation process in a QC consists of a sequence of operations, or logical
gates, in terms of locally tailored Hamiltonians, changing the states of the
qubits through quantum mechanical evolution.  Quantum computation generally
involves logical gates that may affect the state of a single qubit, i.e.
changing $\{\alpha_{\rm in}$,$\beta_{\rm in}\}$ into $\{\alpha_{\rm
out}$,$\beta_{\rm out}\}$, as well as multiple-qubit gates.  The formalism
for quantum information processing is substantially simplified by the
following result proven by Barenco {\it et al} \cite{barenco}: A universal
set of gates, consisting of all one-qubit quantum gates and a single
two-qubit gate, e.g. the controlled-NOT (C-NOT) gate, may be combined to
perform {\em any} logic operation on {\em arbitrarily many} qubits. 
 
The physical realization of qubits begins with demonstration of one-qubit
gates and the C-NOT quantum gate for one and two qubits.  After successfully
performing these basic logic operations at the one and two qubits stage, the
next step is to scale up, eventually achieving a large scale QC of $\sim
10^6$ qubits.
So far, 15 is the largest number for which Shor's
factorization was implemented in a physical system \cite{chuang01}.
This factorization required coherent control over
seven qubits.

Many physical systems have been proposed as
candidates for qubits in a QC, ranging from those in atomic physics, optics,
to those in various branches of condensed matter physics \cite{Reviews}. 
Among the more prominent solid state examples are electron or nuclear spins
in semiconductors \cite{igorrmp,SSReviews}, including electron spin in
semiconductor quantum dots \cite{Exch,Imam} and donor electron or nuclear
spins in semiconductors \cite{Kane,Privman,Vrijen}.  

Silicon donor-based QC schemes are particularly attractive because doped
silicon makes a natural connection between present microelectronic devices
and perspective quantum mechanical devices.  Doping in semiconductors has had
significant technological impact for the past fifty years and is the basis of
current mostly silicon-based microelectronics technology.  As transistors and
integrated circuits decrease in size, the physical properties of the devices
are becoming sensitive to the actual configuration of impurities
\cite{voyles}.  In this context, the first proposal of donor-based silicon
quantum computer (QC) by Kane \cite{Kane}, in which the nuclear spins of the
monovalent $^{31}$P impurities in Si are the qubits, has naturally created
considerable interest in revisiting all aspects of the donor impurity problem
in silicon, particularly in the Si:$^{31}$P system.

In principle, both spin and electronic orbital degrees of freedom can be used
as qubits in semiconductor nanostructures.  A great advantage of orbital (or
equivalently, charge) qubits is that qubit-specific measurements are
relatively simple because measuring single charge states involves
well-developed experimental techniques using single-electron transistors
(SET) or equivalent devices \cite{SCT}.  A major disadvantage of solid state
charge qubits is that these orbital states are highly susceptible to
interactions with the environment that contains all the stray or unintended
charges inevitably present in the device, so that the decoherence time is
generally far too short (typically picoseconds to nanoseconds) for quantum
error correction to be useful.  A related problem is that inter-qubit
coupling, which is necessary for the implementation of two-qubit gate
operations essential for quantum computation, is often the long-range dipolar
coupling for charge qubits.  This makes it difficult to scale up the
architecture, since decoherence grows with the scaling-up as more and more
qubits couple to each other via the long-range dipolar coupling.  However,
the strong interactions make the orbital states an excellent choice for
studying qubit dynamics and qubit coupling in 
the solid state nanostructure environment. 

Spin qubits in semiconductor nanostructures have complementary advantages (and
disadvantages) compared with charge qubits based on quantized orbital states. 
A real disadvantage of spin qubits is that a single electron spin (not to
mention a single nuclear spin) is difficult to measure rapidly, although
there is no fundamental principle against
the measurement of a Bohr magneton.  The great advantage of spin qubits
is the very long spin coherence times, which even for electron spins can be
milliseconds  in silicon at low temperatures.  In addition to the coherence
advantage, spin qubits also have a considerable advantage that the exchange
gate \cite{Exch}, which provides the inter-qubit coupling, is exponentially
short-ranged and nearest-neighbor in nature, thus allowing precise control and
manipulation of two-qubit gates.  There is no fundamental problem arising from
the scaling-up of the QC architecture since exchange interaction couples only
two nearest-neighbor spin qubits independent of the number of qubits.  

We provide here a brief perspective on spin and charge 
qubits in silicon with electron spins or charge states in shallow 
P donor levels in Si being used as qubits.
In Sec.~II we present some background on the classic problem of the 
shallow donor in silicon, describing it through two complementary 
approaches: The effective mass theory and the tight-binding formalism. 
In Sec.~III we analyze the response of the donor electron to an applied 
uniform field, and conclude that electric field control over 
the donor electron does not present additional complications due 
to the Si host electronic structure characteristics. 
Sec.~IV is devoted to the exchange coupling for a 
donor pair in Si, which is highly sensitive to interdonor positioning. 
We review the basic formalism leading to this behavior, and also describe 
attempts to overcome it, namely by considering donors in strained Si,  
and by refining the theoretical formalism for the problem.  
The feasibility of charge qubits based on P$^+_2$ molecular ions 
in Si is investigated in Sec.~V, where we focus on the tunnel coupling and 
charge coherence in terms of electron-phonon coupling.

\section{Single donor in Silicon}

Silicon is a group-IV element, so that when a Si atom 
at a lattice site ${\bf R}_0$ in the bulk is 
replaced by a group-V element like P, the simplest description 
for the electronic behavior of the additional electron is a hydrogenic 
model, in which this electron is subject to the Si crystal 
potential perturbed by a screened Coulomb potential produced by the 
impurity ion:
\begin{equation}
V({\bf r})=-\frac{e^2}{\epsilon|{\bf r}-{\bf R}_0|} \,.
\label{eq:coul}
\end{equation}
The static dielectric constant of Si, $\epsilon = 12.1$, indicates that 
the donor confining potential is weaker than the bare hydrogen atom potential, 
leading to larger effective Bohr radii and smaller binding energies, so that
donors are easily ionized (also known as shallow donors). 

In this section we briefly review basic properties concerning the donor      
ground state wavefunction within two complementary formalisms: 
The effective mass theory (EMT), which is a reciprocal space formalism, and
the 
tight-binding (TB) formalism, which is a real space scheme. 
EMT exploits the duality between real and reciprocal space, 
where delocalization in real space leads to localization in $k$-space.
Since shallow donor wavefunctions are expected to extend over several lattice
constants in real space (the lattice parameter of Si crystal is $a_{\rm
Si}=5.4$ \AA), 
it is written in terms of the bulk eigenstates for 
one or a few $k$-vectors at the lower edge of the conduction band.  
The TB description is a microscopic atomistic formalism, in which the basis
set 
for the donor wavefunction expansion consists of atomic orbitals localized 
at the individual atoms. 

\subsection{Effective mass theory}

The bound donor electron Hamiltonian for an impurity at site ${\bf R}_0$ is
written as
\begin{equation}
{\cal H}_0={\cal H}_{SV}+{\cal H}_{VO} \,.
\label{eq:h0}
\end{equation}
The first term, ${\cal H}_{SV}$, is the single-valley Kohn-Luttinger 
Hamiltonian \cite{Kohn}, which includes the single particle kinetic energy, 
the Si periodic potential, and the screened impurity Coulomb potential 
in Eq.~(\ref{eq:coul}). 
The second term of Eq.~(\ref{eq:h0}), ${\cal H}_{VO}$, includes the 
inter-valley coupling effects due to the presence of the impurity 
potential.

Following the EMT assumptions, the donor electron eigenfunctions are written 
on the basis of the six unperturbed Si band edge Bloch states 
$\phi_\mu = u_\mu(\bf r) e^{i {\bf k}_{\mu}\cdot {\bf r}}$ 
[the conduction band of bulk Si has six degenerate minima
$(\mu=1,\ldots,6)$, located along the $\Gamma-$X axes of the Brillouin zone
at $|{\bf k}_\mu|\sim 0.85(2\pi/a_{\rm Si})$ from the $\Gamma$ point]:
\begin{equation}
\psi_{{\bf R}_0} ({\bf r}) = \frac{1}{\sqrt{6}}\sum_{\mu = 1}^6  F_{\mu}({\bf
r}-
{\bf R}_0) u_\mu({\bf r}) e^{i {\bf k}_{\mu}\cdot ({\bf r}-{\bf R}_0)}\,.
\label{eq:sim}
\end{equation}
In Eq.~(\ref{eq:sim}), $F_{\mu}({\bf r}-{\bf R}_0)$ are envelope functions 
centered at ${\bf R}_0$, for which we adopt the anisotropic Kohn-Luttinger
form, e.g., for $\mu = z$, 
$F_{z}({\bf r}) = \exp\{-[(x^2+y^2)/a^2 + z^2/b^2]^{1/2}\}/\sqrt{\pi a^2 b}$.
The effective Bohr radii $a$ and $b$ are variational parameters chosen to
minimize $E_{SV} = \langle\psi_{{\bf R}_0}| {\cal H}_{SV} |\psi_{{\bf
R}_0}\rangle$, leading to $a=25$ \AA, $b=14$ \AA, in agreement with the
expected increased values with respect to bare atoms.

\begin{figure}
\includegraphics[width=4.1in]
{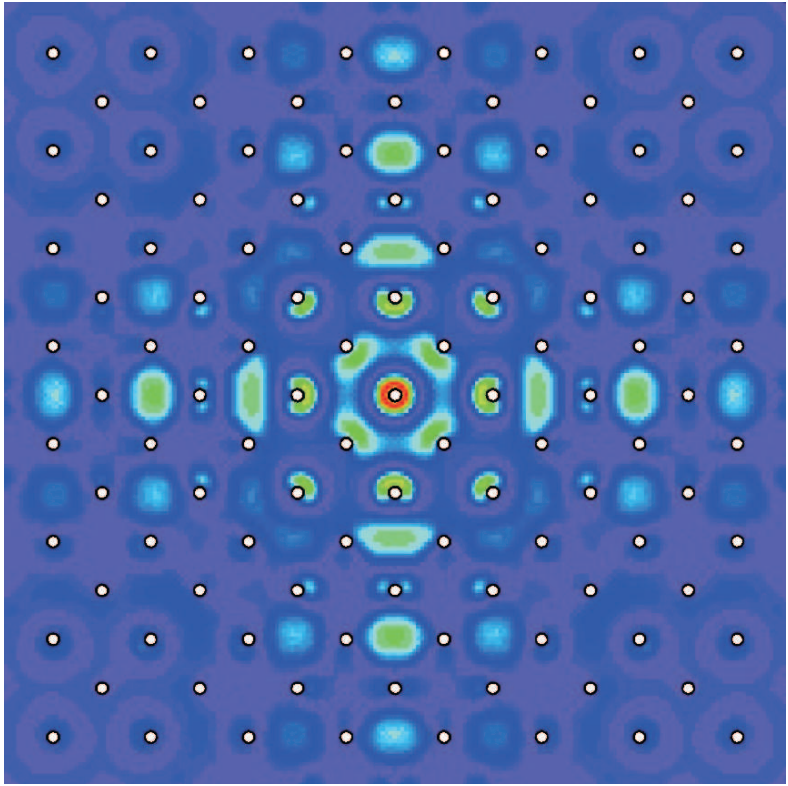}
\protect\caption[donor charge density]
{\sloppy{ (Color)
Electron probability density on the (001) plane of bulk Si 
for the ground state of a donor in Si within the Kohn-Lutttinger 
effective mass theory.
The white dots give the in-plane atomic sites.
}}
\label{charge}
\end{figure}

The ${\cal H}_{SV}$ ground state is six-fold degenerate.  This degeneracy is
lifted by the valley-orbit interactions included here in ${\cal H}_{VO}$, 
leading to the nondegenerate ($A_1$-symmetry) ground state in (\ref{eq:sim}).
Fig.~\ref{charge} gives the charge density $|\psi_{{\bf R}_0} ({\bf r})|^2$
for this 
state, where the periodic part of the conduction band edge Bloch functions
were obtained from 
{\it ab-initio} calculations, as described in Ref.~\onlinecite{KCHD}. 
The impurity site ${\bf R}_0$, corresponding to the higher charge density, 
is at the center of the frame. It is interesting that, except for this central
site, 
regions of high charge concentration and atomic sites do not necessarily
coincide, 
because the charge distribution periodicity imposed by the plane-wave part of
the Bloch 
functions is $2\pi/k_\mu$, incommensurate with the lattice period. 

\subsection{Tight-binding description for P donor in silicon}

The TB Hamiltonian for the impurity problem is written as: 
\begin{equation}
H=\sum\limits_{ij}\sum_{\mu \nu}h_{ij}^{\mu\nu} \,
c_{i\mu}^{\dagger}c_{j\nu}+\sum_
{i,\nu} U({\bf R}_i) \, c_{i\nu}^{\dagger}c_{i\nu} 
\label{hamilton}
\end{equation}
where $i$ and $j$ label the atomic sites, $\mu$ and $\nu$ denote the
atomic orbitals and spins, and $c_{i\nu}^{\dagger},~ c_{i\nu}$ are creation 
and annihilation operators for the atomic states.
We do not include spin-orbit corrections, thus all terms are spin-independent. 
The matrix elements $h_{ij}^{\mu\nu}$ define the on-site energies
and first and second neighbors hopping for the bulk material, for which we
take 
the parametrization given in  Ref.~\onlinecite{Klimeck}. 
The donor impurity potential is included in the perturbation term 
$U({\bf R}_i)$, the same as Eq.~(\ref{eq:coul}), but  
in a discretized form restricted to the lattice sites:
\begin{equation} 
U\left({\bf R}_{i}\right)=-\frac{e^{2}}{\epsilon r_{i}},
\label{potential}
\end{equation}
where $r_{i}$ is the distance of site $i$ to the impurity site. 
At the impurity site $\left(r_{i}=0\right)$, the perturbation potential is 
assigned the value $-U_{0}$, a parameter describing central cell effects 
characteristic of the substitutional species.  
We take $U_0=1.48$ eV, which leads to the experimentally observed 
binding energy  of P in Si, 45.6 meV \cite{martins04}.
Detailed comparison of the TB donor ground state wavefunction with
Kohn-Luttinger 
EMT, performed in Ref.~\onlinecite{martins04}, shows that 
the EMT oscillatory behavior coming from the interference
among the plane-wave part of the six $\phi_\mu$ is well captured by the TB
envelope function. The good agreement between TB and K\&L is limited to
distances from the impurity site larger than a few lattice parameters ($\sim$~1~nm). 
Closer to the impurity, particularly at the impurity site, the TB results become 
considerably larger than the K\&L prediction, in agreement with experiment.

The TB problem is numerically solved by restricting the real-space description  
to a supercell in which periodic boundary conditions are applied.  For the
single donor problem, the supercell is taken to be large enough so that 
convergence in the results is achieved \cite{martins04,martins02}.

\section{Electric-field control of shallow donor in silicon}

Logic operations in quantum computer architectures based on P donors in Si  
involve the response of the bound electron wavefunctions to voltages applied 
to a combination of metal gates separated by a barrier
material (e.g. SiO$_2$) from the Si host. 
The A-gate (according to the nomenclature originally proposed by Kane
\cite{Kane}), 
placed above each donor site, pulls the electron wavefunction away from the
donor, aiming at
partial reduction \cite{Kane} or total cancellation \cite{Skinner} of the
electron-nuclear hyperfine coupling in architectures where the qubits are the
$^{31}$P nuclear spins. In a related proposal based on the donor electron
spins as qubits \cite{Vrijen}, the gates drive the electron wavefunction into
regions of different $g$-factors, allowing the exchange coupling between
neighboring electrons to be tuned.

\begin{figure}
\includegraphics[height=4in]
{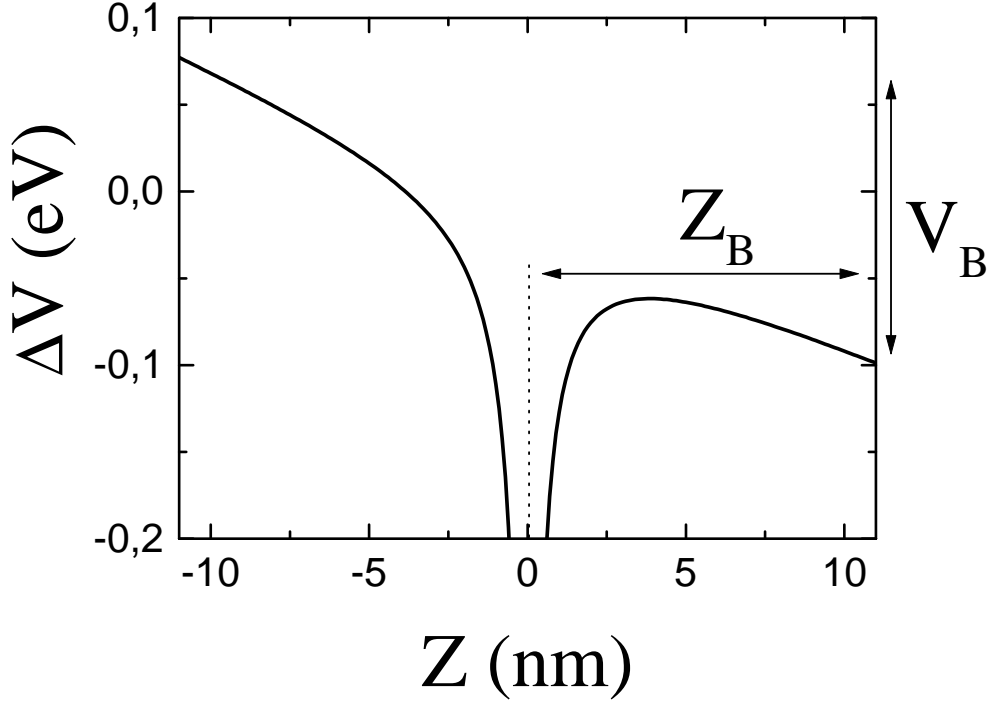}
\protect\caption[inset in fig3 from martins04]
{\sloppy{
Schematic representation of the perturbation
potential along the z-axis to be added to the bulk Si Hamiltonian due 
to the impurity at ${\bf R}=0$
and to a uniform electric field in the negative $z$ direction. 
This particular plot corresponds numerically to a supercell length of 
$L_z = 40 a_{\rm Si}$ and to an electric field of 80 kV/cm.
}}
\label{inset}
\end{figure}

We present here a simplified model of the A-gate operation by considering the 
Si:P system under a uniform electric field and near a barrier. 
Following Ref.~\onlinecite{martins04}, we describe the electronic 
problem within the TB approach, where the basic Hamiltonian is given in 
Eq.~(\ref{hamilton}), with the perturbation term including the Coulomb
potential 
as in Eq.~(\ref{potential}), plus the contribution of  
a constant electric field of amplitude $E$ applied along the 
$[00\bar1]$ direction:
\begin{equation} 
U({\bf R}_i) = -\frac{e^{2}}{\epsilon r_{i}} - |e| E z_i.
\label{electric}
\end{equation}
The overall preturbation potential along the z-axis is represented in Fig.~\ref{inset}. 
We take the origin of the potential at the impurity site, ${\bf R}_0$, at 
the center of the supercell. 
Periodic boundary conditions lead to a discontinuity in the
potential at the supercell boundary $z_i=Z_B$, where $Z_B$ is half of the
supercell length along [001] or, equivalently, the distance from the
impurity to the Si/barrier interface. The potential discontinuity,
$V_B=2|e|EZ_B$,
actually has a physical meaning in the present study: It models the potential
due to the barrier material layer above the Si host (see Fig.~\ref{inset}).

\begin{figure}
\includegraphics[height=4in]
{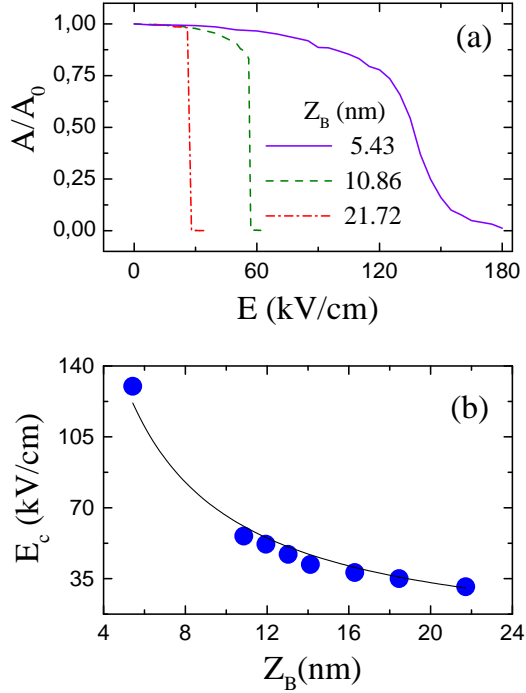}
\protect\caption[fig3 from martins04]
{\sloppy{
(a) TB envelope function squared at the impurity site
under applied field $E$, normalized to the zero-field value, for the indicated
values of the impurity-Si/barrier interface distance $Z_B$. (b) Dependence
of the critical field $E_c$ on $Z_B$. The solid line is a best fit of the form
$E_c \propto 1/Z_B$. 
}}
\label{hyperfine}
\end{figure}

A description of the A-gate operations may be inferred
from the behavior of the TB envelope function squared (this function is defined at each
lattice site as the sum of the squared TB wavefunction expansion coefficients at this site) 
at the impurity site under applied field $E$, normalized to the zero-field value:
\begin{equation}
A/A_0 = |\Psi_{TB}^E({\bf R}_0)|^2/|\Psi_{TB}^{E=0}({\bf R}_0)|^2.
\label{ratio}
\end{equation}
The notation here indicates that this ratio should follow a behavior similar
to that for the hyperfine coupling constants between the donor nucleus and
electron with $(A)$ and without $(A_0)$ external field. 
The ratio in (\ref{ratio}) is plotted in
Fig.~\ref{hyperfine}(a) for three values of the impurity depth with respect to
the
Si/barrier interface. Calculations for $Z_B$=10.86~nm were performed with
cubic supercells $(L=40$\,$a_{\rm Si}$), while for $Z_B$= 5.43 and 21.72~nm
tetragonal
supercells with $L_x=L_y=40$\,$a_{\rm Si}$ and $L_z=20$\,$a_{\rm Si}$ and
80\,$a_{\rm Si}$ respectively were used.
At small field values we obtain a quadratic decay of $A/A_0$ with $E$,
in agreement with the perturbation theory results for the hydrogen atom.
At large enough fields, $|\Psi_{TB}^E({\bf R}_0)|^2$ becomes vanishingly
small, and the transition between the two regimes is qualitatively different
according to $Z_B$: For the largest values of $Z_B$ we get an abrupt
transition
at a critical field $E_c$, while smaller $Z_B$ (e.g. $Z_B=5.43$~nm) lead to a
smooth decay, similar to the one depicted in Ref.~\onlinecite{Kane}. 
In this latter case, we define $E_c$ as the field for which the curve 
$A/A_0$ vs $E$ has an inflection point, where $A/A_0\sim 0.5$, 
thus $E_c(5.43 \rm{nm}) = 130$ kV/cm. 
We find that the decrease of $E_c$ with $Z_B$ follows a simple rule
$E_c \propto 1/Z_B$, as given by the solid line in Fig.~\ref{hyperfine}(b).

\begin{figure}
\includegraphics[height=4in]
{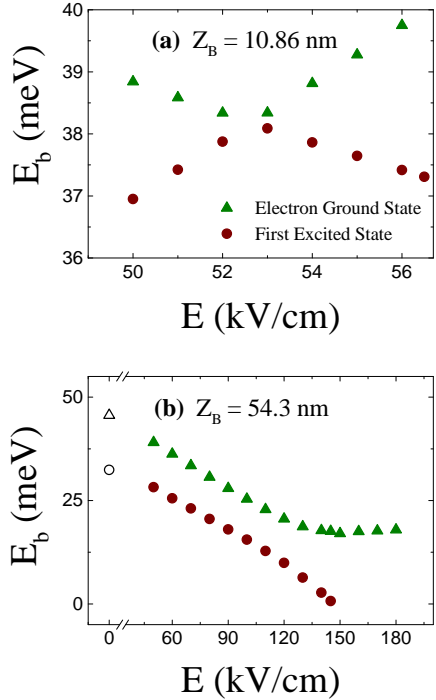}
\caption{Calculated binding energies versus electric field intensity
of the two lowest donor electron states. (a) For $Z_{B}=10.86$~nm
the energies reveal a crossing regime. (b) Anticrossing of the
two lowest electron states for $Z_{B}=5.43$~nm. The open symbols correspond
the zero field calculated values: 45.6 meV and 32.4 meV, in good 
agreement with experiment.
\label{binding}}
\end{figure}

The above results may be understood within a simple picture of the electron
in a double well potential, the first well being most attractive at the
impurity site, $V(\mathbf{R_0}=0)=-U_0$, and the second well at the barrier
interface, $V(z=Z_B)= -V_B/2=-|e|EZ_B$ neglecting the Coulomb potential
contribution at the interface. 
An internal barrier separates the two wells and, for a fixed $E$, 
this internal barrier height and width increase with $Z_B$. 
Deep donor positioning leads to a weaker coupling between
the states localized at each well, even close to level degeneracy, resulting
the
level crossing behavior of the two lowest donor-electron states 
illustrated in Fig.\ref{binding}(a). 
For a donor positioned closer to the interface, the internal barrier gets
weaker, 
enhancing the coupling between levels localized in each well and 
leading to wavefunction superposition and to the anticrossing behavior
illustrated in 
Fig.\ref{binding}(b). 
The scaling of $E_c$ with $1/Z_B$ may also be understood assuming that the
critical field
corresponds to the crossing of the ground state energies of two wells: The
Coulomb potential well and an approximately triangular well at the barrier.
Since the relative depths of the wells increases with $EZ_B$, and assuming
that the ground states energies are fixed with respect to each well's depth,
the $E_c\propto 1/Z_B$ behavior naturally results.

The minimum gap at the anticrossing in 
Fig.~\ref{binding}(b) is $\simeq 9.8$\,meV, which allows for 
adiabatic control of the electron by the A-gate within switching times of the 
order of picoseconds, as discussed in Ref.~\cite{martins04}. 
This is a perfectly acceptable time for the operation of
A-gates in spin-based Si QC, given the relatively long electron spin coherence
times (of the order of a few ms) in Si.

We remark that the Bloch phases interference behavior in the donor
wavefunctions  
are well captured in the  TB wavefunctions, and that the results above 
demonstrate that electric field control over single donor wavefunctions, 
such as proposed in A-gate operations, \cite{Kane,Vrijen,Skinner} do not
present 
additional complications due to the Si band structure. 
The only critical parameter is the donor positioning below the Si/barrier
interface, 
which should be chosen and controlled according to physical criteria such as
those 
discussed here.

\section{Donor Electron Exchange in Silicon}

An important issue in the study of donor-based Si QC architecture is coherent
manipulations of spin states as required for the quantum gate operations.  
In particular, 
two-qubit operations, which are required for a universal QC,
involve precise control over electron-electron exchange
\cite{Exch,Kane,Vrijen,HD}.  
Such control can presumably be achieved by fabrication
of donor arrays with well-controlled positioning and surface gate potential 
\cite{Obrien,encapsulation,implant,schenkel03}.  
However, electron exchange in bulk silicon has spatial
oscillations \cite{Andres,KHD1} on the atomic scale due to valley 
interference arising from the particular six-fold degeneracy of the bulk Si
conduction band.  These exchange oscillations place heavy burdens on device
fabrication and coherent control \cite{KHD1}, because of the very high
accuracy and tolerance requirements for placing each donor inside the Si unit
cell, and/or for controlling the external gate voltages.  

The potentially severe consequences of the exchange-oscillation problem for
exchange-based Si QC architecture motivated us and other researchers to
perform theoretical studies with increasingly sophisticated formalisms,
incorporating perturbation effects due to applied strain\cite{KHD2} or gate
fields \cite{wellard03}.  These studies, all performed within the standard
Heitler-London (HL) formalism \cite{slater}, essentially reconfirm the
originally reported difficulties \cite{KHD1} regarding the sensitivity of the
electron
exchange coupling to precise atomic-level donor positioning, indicating that
they may not be completely overcome by applying strain or electric fields. 
The sensitivity of the calculated exchange coupling to donor relative 
position originates from interference between the plane-wave parts of the six
degenerate Bloch states associated with the Si conduction-band minima. More
recently \cite{KCHD} we have assessed the robustness of the HL approximation
for the two-electron donor-pair states by relaxing the phase pinning at donor
sites.

In this section, we first review the main results regarding exchange
coupling for a donor pair in relaxed bulk Si, and its high sensitivity to 
interdonor positioning.  
We then discuss ways to overcome this behavior, namely considering donors in 
strained Si and the more general {\em floating-phase} HL formalism. 
We show that strain may partially alleviate the exchange oscillatory behavior, 
but it cannot entirely overcome the problem. 
From the {\em floating-phase} HL approach results, 
our main conclusion is that, for all practical purposes, the
previously adopted HL wavefunctions are robust, and the exchange 
sensitivity to donor positioning obtained in 
Refs.~\onlinecite{KHD1,KHD2,wellard03} persists 
in the more sophisticated theory of Ref.~\onlinecite{KCHD}.

\subsection{Donor Electron Exchange in Relaxed Bulk Silicon}

The HL approximation is a reliable scheme to calculate electron exchange for a
well-separated pair of donors (interdonor distance much larger than the donor
Bohr radii)
\cite{slater}. Within HL, the lowest energy singlet and triplet wavefunctions
for two electrons bound to a donor pair at sites $\mathbf{R}_A$ and
$\mathbf{R}_B$, are written as properly symmetrized and normalized
combinations of
$\psi_{\mathbf{R}_A}$ and $\psi_{\mathbf{R}_B}$ [as defined in
Eq.(\ref{eq:sim})]
\begin{equation}
\Psi^s_t({\mathbf r}_1,{\mathbf r}_2) = \frac{1}{\sqrt{2(1\pm S^2)}}
\left[ \psi_{{\bf R}_A}({\bf r}_1) \psi_{{\bf R}_B}({\bf r}_2) \pm 
\psi_{{\bf R}_B}({\bf r}_1) \psi_{{\bf R}_A}({\bf r}_2) \right],
\label{eq:hl}
\end{equation}
where $S$ is the overlap integral and the upper (lower) sign corresponds to
the singlet (triplet) state.  The energy expectation values for these states, 
$E^s_t = \langle\Psi^s_t|{\cal H}|\Psi^s_t\rangle$, give the exchange
splitting through their difference, $J=E_t-E_s$.  We have previously derived
the expression for the donor electron exchange splitting \cite{KHD2,KCHD},
which we reproduce here:
\begin{eqnarray} 
J({\bf R}) = \frac{1}{36}\sum_{\mu, \nu} {\cal J}_{\mu \nu}
({\bf R}) \cos ({\bf k}_{\mu}-{\bf k}_{\nu})\cdot {\bf R}\,,
\label{eq:exch}
\end{eqnarray}
where $\mathbf{R} = \mathbf{R}_A - \mathbf{R}_B$ is the interdonor position
vector and ${\cal J}_{\mu \nu} ({\bf R})$ are kernels determined by the
envelopes and are slowly varying functions of $\mathbf{R}$ \cite{KHD1,KHD2}.  
Note that Eq.~(\ref{eq:exch}) does not involve any oscillatory contribution
from 
$u_{\mu}({\bf r})$, the periodic part of the Bloch functions
\cite{wellard03,KCHD}.  The physical reason for that is clear from
(\ref{eq:sim}): While the plane-wave phases of the Bloch functions are pinned
to the donor sites, leading to the cosine factors in (\ref{eq:exch}), 
the periodic functions $u_\mu$ are pinned to the lattice, regardless of the
donor location. 

\begin{figure}
\includegraphics[width=4in]
{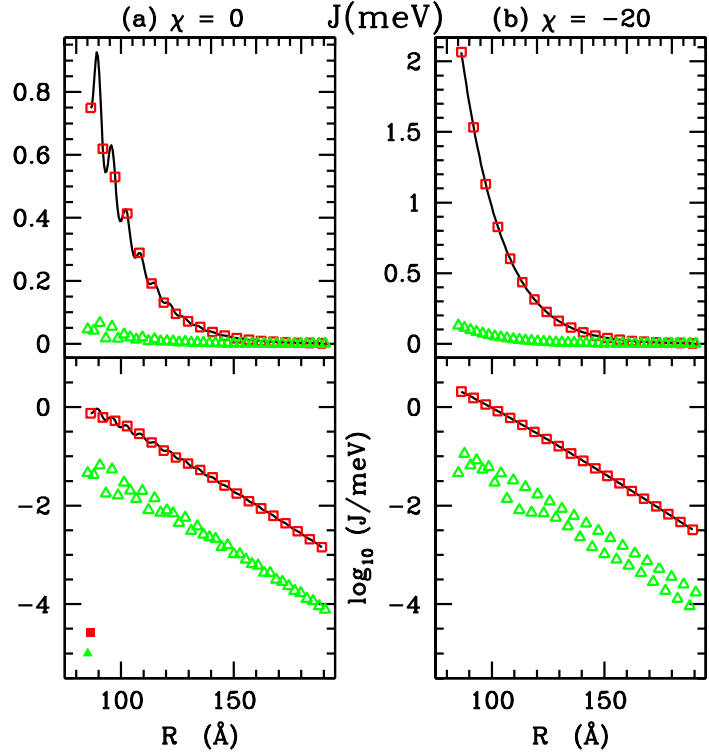}
\protect\caption[exchange with nn shifts]
{\sloppy{
Calculated exchange coupling for a donor pair  versus interdonor distance in
(a) unstrained and (b) uniaxially strained (along $z$) Si.
The open squares correspond to substitutional donors placed exactly along the
[100] axis,  the lines give the calculated values for continuously
varied interdonor distance along this axis, assuming the envelopes
do not change. The open triangles give the exchange for a substitutional
pair {\it almost} along [100], but with one of the donors displaced
by $d_{NN}\sim 2.3$~\AA~ into a nearest-neighbor site.
The lower frames give the same data in a logarithmic scale.
When the floating-phase HL approach is adopted, the results change negligibly; 
the filled symbols on the lower left frame 
give examples of calculated corrections (see text).
}}
\label{fig:nearest}
\end{figure}

As an example of the consequences of the sensitivity of exchange to interdonor 
relative positioning, we present in Fig.~\ref{fig:nearest}(a) 
a case of practical concern involving unintentional donor displacements 
into nearest-neighbor sites, when the two donors belong to different fcc
sublattices.  
The open squares in Fig.~\ref{fig:nearest}(a) give $J({\bf R})$ for
substitutional donors along
the [100] axis, while the open triangles illustrate the different-sublattice
positioning situation, namely ${\bf R} = {\bf R_0} + {\vec \delta}_{NN}$ with
${\bf R_0}$ along the [100] axis and ${\vec \delta}_{NN}$ ranging over the
four nearest-neighbors of each ${\bf R_0}$
($d_{NN}=|\vec\delta_{NN}|= a_{\rm Si} \sqrt{3}/4 \sim 2.34$~\AA).  The lower
panel 
of the figure presents the same data on a logarithmic scale, showing that
nearest-neighbor displacements lead to an exchange coupling reduction by one
order of magnitude when compared to $J({\bf R_0})$.

\subsection{Strained Si}

The extreme sensitivity of $J({\bf
R})$ to interdonor positioning can be eliminated for on-lattice substitutional
impurities in uniaxially strained Si (e.g. along the $z$ axis) commensurately
grown over Si-Ge alloys {\it if inter-donor separation {$\bf R$} remains
parallel to the
interface $x$-$y$ plane} \cite{KHD2}.  The strain is accommodated in the Si
layer by
increasing the bond-length components parallel to the interface and
decreasing those along $z$, breaking the cubic symmetry of the lattice and
lowering the six-fold degeneracy of the conduction band minimum to two-fold.
In this case, the valley populations in the donor electron
ground state wave function in Eq.~(\ref{eq:sim}) are not all equal to
$1/\sqrt{6}$, 
but are determined from a scalar valley strain parameter $\chi$, 
which quantifies the amount of strain.
Fig.~\ref{fig:nearest}(b) gives $J({\bf R})$ in uniaxially strained (along
$z$ direction) Si for $\chi=-20$ (corresponding to Si grown over a Si-Ge alloy 
with 20\% Ge-content) for the same relative positioning of the
donor pairs as in Fig.~\ref{fig:nearest}(a).  Notice that the exchange
coupling is enhanced by about a factor of 2 with respect to the relaxed Si
host, but the order-of-magnitude reduction in $J$ caused by displacements of
amplitude $d_{NN}$ into nearest-neighbor sites still persists as ${\vec
\delta}_{NN}$ is not parallel to the $x$-$y$ plane.

\subsection{Floating-phase Heitler-London approach}

In Refs.~\onlinecite{KHD1} and \onlinecite{KHD2}, as in the standard HL
formalism presented in subsection IV-A, it is implicitly assumed that the
phases $e^{-i
{\bf k}_{\mu}\cdot {\bf R}_0}$ in Eq.~(\ref{eq:sim}) remain pinned to the
respective donor sites $\mathbf{R}_0 = \mathbf{R}_A$ and $\mathbf{R}_B$,
as we adopt single donor wavefunctions to build the two-electron wavefunction.
Although phase pinning to the donor substitutional site is required for the
ground state of an isolated donor ($A_1$ symmetry) in order to minimize
single electron energy, this is not the case for the lower-symmetry problem
of the donor pair.  In order to minimize the energy of the two-donor system,
here we allow the phases to shift by an amount $\mathbf{\delta R}$ along the
direction of the interdonor vector $\mathbf{R} = \mathbf{R}_B-\mathbf{R}_A$,
so that the single-particle wavefunctions in Eq.~(\ref{eq:hl}) become
\begin{equation}
\psi_{{\bf R}_A} ({\bf r}) = \frac{1}{\sqrt{6}} \sum_{\mu = 1}^6 F_{\mu}({\bf
r} - {\bf R}_A) u_\mu(\bf r) e^{i {\bf k}_{\mu}\cdot ({\bf r}-{\bf R}_A +
{\mathbf \delta R})}
\label{eq:sima}
\end{equation}
and
\begin{equation}
\psi_{{\bf R}_B} ({\bf r}) = \frac{1}{\sqrt{6}} \sum_{\mu = 1}^6 F_{\mu}({\bf
r} - {\bf R}_B) u_\mu(\bf r) e^{i {\bf k}_{\mu}\cdot ({\bf r}-{\bf R}_B -
{\mathbf \delta R})}\,.
\label{eq:simb}
\end{equation}
We take $\mathbf{\delta R}$ as a variational parameter to minimize
$E_s$ and $E_t$. 
Since the phases in Eq.(\ref{eq:sim}) are
responsible for the sensitivity of the exchange coupling to donor 
positioning in Si, this more general variational treatment might lead to
changes in the previously reported  \cite{KHD1,KHD2,wellard03}  behavior of
the two-donor exchange splitting $J=E_t-E_s$.

Minimization of the total energy for the
particular geometry where the donor pair is 87~\AA~ apart along the [100]
direction leads 
the singlet energy decrease of 270~neV, and the triplet energy decrease of
6~neV.
This results in an increase in $J$ by $264$~neV, given by the solid square
in the lower left hand side frame of Fig.~\ref{fig:nearest}.  The floating
phases variational scheme leads to a reduction in both singlet and triplet
states energy, therefore the net variation in $J$ is positive (negative) if
the triplet energy reduction is smaller (larger) than the singlet.  The solid
triangle in Fig.~\ref{fig:nearest} corresponds to a case of negative
variation, obtained when one of the donors in the above geometry is displaced
into a nearest-neighbor site.  Note that the corrections are more than three
orders of magnitude smaller than the calculated $J$ within standard HL.  
In other words, for all practical purposes the fixed-phase
standard HL approximation is entirely adequate for the range of interdonor
distances of interest for QC applications.

From the perspective of current QC fabrication efforts, $\sim 1$ nm accuracy
in single P atom positioning has been recently demonstrated
\cite{encapsulation}, representing a major step towards the goal of obtaining
a regular donor array embedded in single crystal Si.  Exchange
coupling distributions consistent with such accuracy are presented in
Ref.~\onlinecite{KH}, indicating that even such small deviations ($\sim$ 1 nm)
in the
relative position of donor pairs can still lead to significant changes in the
exchange coupling, favoring $J\sim 0$ values.  Severe limitations in
controlling $J$ would come from ``hops'' into different substitutional
lattice sites.  Therefore, precisely controlling of exchange gates in Si
remains an open (and severe) challenge.
As suggested in Ref.~\onlinecite{KHDD}, spatially resolved micro-Raman
spetroscopy 
might provide a valuable diagnostic tool to  characterize local values of
exchange 
coupling between individual spin qubits.

\section{Charge qubits in silicon}

Successful coherent manipulation of electron {\it orbital} states in GaAs has
been achieved for electrons bound to donor impurities \cite{Cole00} as well
as electrons in double quantum dots \cite{Hayashi}.  There were also
suggestions of directly using electron orbital states in Si as the
building blocks for quantum information processing
\cite{hollenberg1,hollenberg2}.  
Specifically, a pair of phosphorus donors that sit relatively close to 
each other (so as to have sizable wave function overlap) form an effective 
hydrogen molecule in Si host material.
Charge qubits may be defined by ionizing one of the bound electrons, thus
leading to a double well potential filled with a single electron: The single
electron ground state manifold, whether it is the two states localized in
each of the wells or their symmetric and anti-symmetric combinations, can
then be used as the two-level system forming a charge
qubit \cite{Ekert,Tanamoto}.  The advantage of such a charge qubit is that it
is easy to manipulate and detect, while its disadvantage, as already
mentioned above, is the generally fast charge decoherence as compared to
spin.

In this section we discuss the feasibility of the P$_2^+$ charge qubit in Si,
focusing on
single qubit properties in terms of the tunnel coupling between the two
phosphorus donors, and charge decoherence of this system in terms of
electron-phonon coupling.
We take into consideration the multi-valley
structure of the Si conduction band and explore whether valley interference
could lead to potential problems or advantages with the operations of P$_2^+$
charge qubits, such as difficulties in the control of tunnel coupling similar
to the control of exchange in two-electron systems discussed in Sec.~IV, or
favorable decoherence properties through vanishing electron-phonon coupling.

\subsection{The P$_2^+$ molecule in Silicon}

We study the simple situation where a single electron is shared by a donor
pair, constituting a P$_2^+$ molecule in Si.  The charge qubit here consists
of the two lowest energy orbital states of an ionized P$_2$ molecule in Si
with only one valence electron in the outermost shell shared by the two P
atoms.  The key issue to be examined is the tunnel coupling and the resulting
coherent superposition of one-electron states, rather than the entanglement
among electrons, as occurs for an exchange-coupled pair of electrons.

The donors are at substitutional sites ${\bf R}_A$ and ${\bf R}_B$ in an
otherwise perfect Si structure.  In the absence of an external bias, 
we write the eigenstates for the two lowest-energy states as a superposition 
of single-donor ground state wavefunctions [as given in Eq.~(\ref{eq:sim})]   
localized at each donor, $\psi_A ({\bf r})$ and $\psi_B ({\bf r})$, similar to
the
standard approximation for the H$_2^+$ molecular ion \cite{slater}.  The
symmetry of the molecule leads to two eigenstates on this basis, namely the
symmetric and antisymmetric superpositions
\begin{equation}
\Psi_\pm({\bf r})=\frac{\psi_A({\bf r}) \pm \psi_B({\bf r})}{\sqrt{2(1 \pm
S)}}~.
\label{eq:wav}
\end{equation}
As described in Ref.~\onlinecite{Hu04}, the energy gap between these 
two states may be written as 
\begin{equation}
\Delta_{\rm S-AS} = 
\frac{2}{1-S^2}\sum_{\mu=1}^6{\delta}_{\mu}({\bf R}) \cos ({\bf k}_{\mu}\cdot
{\bf R})~,
\label{eq:delta}
\end{equation}
where $S$ is the overlap integral between $\psi_A ({\bf r})$ and $\psi_B ({\bf
r})$. 
For ${\bf R} = {\bf R}_A-{\bf R}_B\gg a,b$, the amplitudes $\delta_\mu({\bf
R})$ 
are monotonically decaying functions of the interdonor distance ${\bf R}$, and
$S\ll 1$.  
The dependence of $\delta_\mu$ on $|{\bf R}|$ is qualitatively similar to the
symmetric-antisymmetric gap in the  $H_2^+$ molecule, namely an exponential
decay with power-law prefactors.  The main difference here comes from the
cosine factors, which are related to the oscillatory behavior  of
the donor wavefunction in Si arising from the Si conduction band valley
degeneracy, and to the presence of two pinning centers.

\begin{figure}
\includegraphics[width=4.1in]
{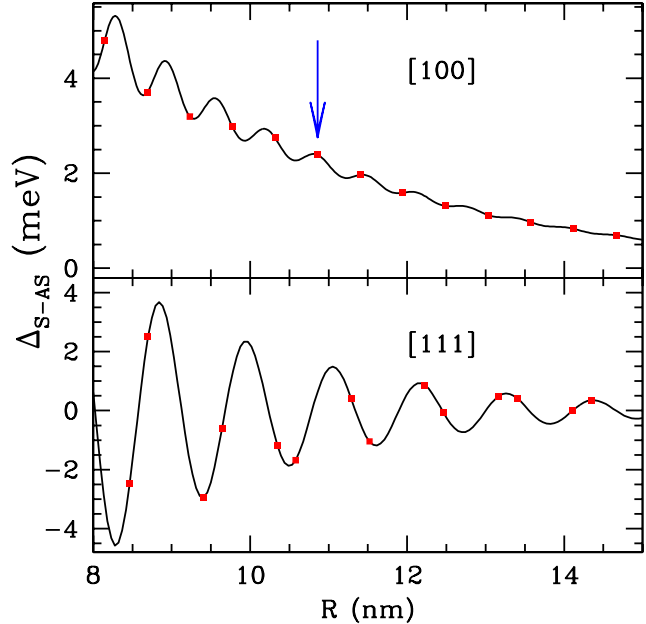}
\protect\caption[two frames with Delta vs R]
{\sloppy{
Symmetric-antisymmetric gap for the $P_2^+$ molecular ion in Si
for the donor pair along the indicated lattice directions. The arrow in the
upper frame indicates the {\it target} configuration analyzed in
Fig.~\ref{fig:distribution}.}}
\label{fig:delta}
\end{figure}

Fig.~\ref{fig:delta} shows the calculated gaps as a function of ${\bf R}$
for a donor pair along two high-symmetry crystal directions.
Two points are worth emphasizing here, which are manifestly different from the
corresponding hydrogenic molecular ion behavior: (i) $\Delta_{S-AS}$ is an
anisotropic and fast oscillatory function of ${\bf R}$; (ii) the sign of
$\Delta_{S-AS}$ may be positive or negative depending on the precise value of
${\bf R}$.  The characteristics mentioned in point (i) are similar to the
exchange coupling behavior previously discussed for the two-electrons neutral
donor pair.\cite{KHD1,KHD2,KCHD}  Point (ii) implies that the $P_2^+$
molecular ion ground state in Si may be symmetric (as in the $H_2^+$
molecular ion case) or antisymmetric depending on the separation between the
two P atoms.  Note that for the two-electron case, the ground state is always
a singlet (i.e. a symmetric two-particle spatial part of the wavefunction
with the spin part being antisymmetric), implying that the exchange $J$ is
always positive for a two-electron molecule.  For a one-electron ionized
molecule, however, the ground state spatial wavefunction can be either
symmetric or antisymmetric.
\begin{figure}
\includegraphics[width=4.1in]
{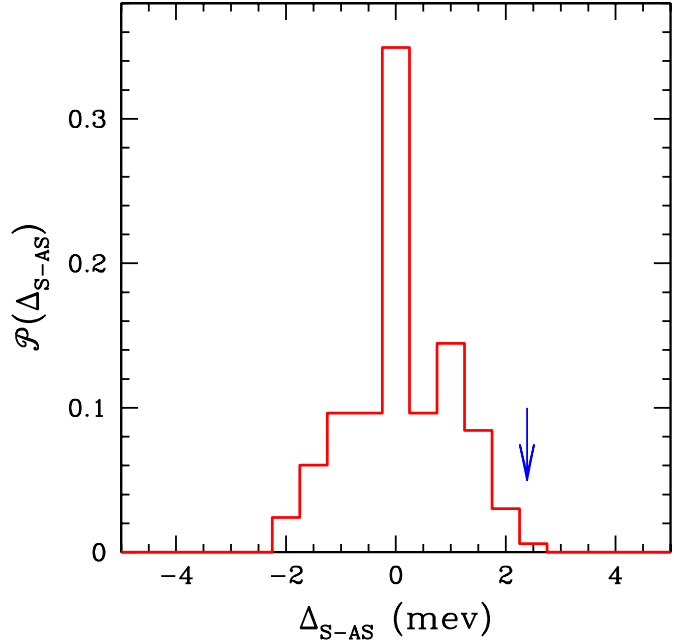}
\protect\caption[distributions of Delta]
{\sloppy{
Probability distribution of the symmetric-antisymmetric gap
for the $P_2^+$ molecular ion in Si.  Donor pairs are approximately aligned
along [100], but with an uncertainty radius $R_u=1$nm with respect to this {\it
target} axial alignment (see text).  The arrow indicates the
gap value for the {\it target} configuration, for which the uncertainty
radius is $R_u=0$.  Notice that the distribution is peaked at
$\Delta_{S-AS} =0$, and not at the {\it target} gap value.
}}
\label{fig:distribution}
\end{figure}

Fig.~\ref{fig:distribution} shows the normalized probability distribution
for the $\Delta_{S-AS}$ gap values when the first donor is kept fixed at ${\bf
R}_A$ and the second donor is placed at a site 20 lattice parameters away
($\sim$ 108.6 \AA), along the [100] axis.  This {\it target} configuration is
indicated by an arrow in Fig.~\ref{fig:delta}.  We allow the second donor
position ${\bf R}_B$ to visit all possible substitutional diamond lattice
positions within a sphere of radius $R_u$ centered at the attempted position.
Our motivation here is to simulate the realistic fabrication of a P$_2^+$
molecular ion with fixed inter-atomic distance in Si with the state of the
art Si technology, in which there will always be a small ($R_u \sim 1-3$ nm)
uncertainty in the precise positioning of the substitutional donor atom
within the Si unit cell.  We would like to estimate the resultant randomness
or uncertainty in $\Delta_{S-AS}$ arising from this uncertainty in ${\bf
R}_B$.  For $R_u = 0$, i.e., for ${\bf R} = 20 a_{\rm Si} \hat x$,
$\Delta_{S-AS}
\simeq 2.4$ meV, given by the arrows in Fig.~\ref{fig:distribution}.  We
incorporate the effect of small uncertainties by taking $R_u = 1$nm,
corresponding to the best reported degree of accuracy in single P atom
positioning in Si \cite{encapsulation}.  These small deviations completely
change the qubit gap distribution, as given by the histogram in
Fig.~\ref{fig:distribution}, strongly peaked around zero.   
Further increasing $R_u$ leads to broader distributions of the gap values, 
though still peaked at zero \cite{Hu04}.  This broadening is due to the fast
increase in the number of lattice sites inside the sphere of radius $R_u$,
thus contributing to the distribution, as $R_u$ increases.  We conclude that
the valley interference between the six Bloch states leads to a strong
suppression of the qubit fidelity since the most probable $\Delta_{S-AS}$
tends to be zero.

A very small $\Delta_{S-AS}$  is undesirable in defining the two states
$|0\rangle$ and $|1\rangle$ forming the charge qubit.  If we take them to be
the symmetric and anti-symmetric states given in Eq.~(\ref{eq:wav}), the fact
that they are essentially degenerate means that, when one attempts to
initialize the qubit state at $|0\rangle$, a different combination
$\alpha|0\rangle+\beta|1\rangle$ might result.  Well defined qubits may still
be defined under a suitable applied external bias, so that the electron
ground state wavefunction is localized around one of the donors, say at
lattice site ${\bf R}_A$, and the first excited state is localized around
${\bf R}_B$.

Single qubit rotations, used to implement universal quantum
gates \cite{nielsen00}, might in principle be achieved by adiabatic
tunneling of the electron among the two sites under controlled axially
aligned electric fields through bias sweeps \cite{Barrett}.  When, at zero
bias, the ground state is {\it not} well separated by a gap from the first
excited state, severe limitations are expected in the adiabatic manipulation
of the electron by applied external fields.  In other words, the fidelity of
the single qubit system defining the quantum two-level dynamics will be
severely compromised by the valley interference effect.

\subsection{Electron-phonon coupling}

Two key decoherence channels for charge qubits in solids are background charge
fluctuations and electron-phonon coupling \cite{Hayashi}.  The former is
closely related to the sample quality (e.g., existences of stray charges and
charged defects in the system) and is extrinsic, while the latter is
intrinsic.  Here we focus on the electron-phonon coupling.  A critical
question for the P$_2^+$ molecular ion in Si is whether the Si bandstructure
and the associated charge density oscillations \cite{KCHD} lead to any
significant modification of the electron-phonon coupling matrix elements. 
The relevant terms for the electron-phonon interaction in Si takes the form: 
\begin{equation}
H_{ep} = D \sum_{\bf q} \left( \frac{\hbar}{2 \rho_m V \omega_{\bf q}}
\right)^{1/2} |{\bf q}| \rho({\bf q}) (a_{\bf q}+a_{-\bf q}^{\dagger}) \,,
\end{equation}
where $D$ is the deformation constant, $\rho_m$ is the mass density of the
host material, $V$ is the volume of the sample, $a_{\bf q}$ and $a_{-\bf
q}^\dagger$ are phonon annihilation and creation operators, and $\rho({\bf
q})$ is the Fourier transform of the electron density operator. 
For the two-donor situation, where we are only interested in the two lowest
energy single-electron eigenstates,  the
electron-phonon coupling Hamiltonian is conveniently written in this
quasi-two-level basis in terms of the Pauli spin matrices $\sigma_x$ and
$\sigma_z$ (where spin up and down states refer to the two electronic
eigenstates, labeled $\{|+\rangle,|-\rangle\}$):
\begin{eqnarray}
H_{ep} & = & D \sum_{\bf q} \left( \frac{\hbar}{2 \rho_m V \omega_{\bf q}}
\right)^{1/2} |{\bf q}| \left(A_r \sigma_x + A_{\varphi} \sigma_z \right)
\left( a_{\bf q}+a_{-\bf q}^{\dagger} \right) \,, \nonumber \\
A_r & = & \langle -|e^{i{\bf q}\cdot{\bf r}}|+\rangle \,, \nonumber \\
A_{\varphi} & = & \frac{1}{2} \left( \langle +|e^{i{\bf q}\cdot{\bf
r}}|+\rangle -
\langle -|e^{i{\bf q}\cdot{\bf r}}|-\rangle \right) \,.
\label{eq:el-ph}
\end{eqnarray}
Here the term proportional to $\sigma_x$ can lead to transition between the
two electronic eigenstates and is related to relaxation; while the term
proportional to $\sigma_z$ only causes energy renormalization of the two
electronic levels, but no state mixing, so that it only leads to pure
dephasing for the electronic charge states.

Calculations of the matrix elements involved in Eq.~(\ref{eq:el-ph}), 
reported in Ref.~\onlinecite{Hu04}, lead to the conclusion that 
the electron-phonon coupling for a P$_2^+$ molecular ion
in Si formally behaves very similarly to that for a single electron
trapped in a GaAs double quantum dot.  For example, the relaxation matrix
element is proportional to
\begin{eqnarray}
A_r & = & (ab^* - a^*b e^{i {\bf q} \cdot {\bf R}}) \int d{\bf r} ~ e^{i{\bf
q} \cdot {\bf r}} [\varphi ({\bf r})]^2 \nonumber \\
& & + (|b|^2 - |a|^2) \int d{\bf r} ~ e^{i{\bf q} \cdot {\bf r}} \varphi ({\bf
r}) \varphi ({\bf r-R}) \,,
\end{eqnarray}
where the more complicated multi-valley bandstructure of Si and the
strong inter-valley coupling introduced by the phosphorus donor atoms only
strongly affect the off-site (thus small) contribution to the electron-phonon
coupling, so that they do not cause significant changes in the overall
electron-phonon coupling matrix elements.  Therefore, available
estimates\cite{Barrett,Fedichkin} of decoherence induced by electron-phonon
coupling based on a single-valley hydrogenic approximation in the P$_2^+$
system in Si should be valid.  In other words, the multi-valley quantum
interference effect does not provide any particular advantage (or
disadvantage) for single qubit decoherence in the Si:P donor charge-based
QC architecture.

\section{Summary}

In summary, we have briefly reviewed physical aspects related to some of the  
relevant building blocks for the implementation of donor spin and charge
qubits in silicon: Electric field control of a single donor, the exchange
gate for two spin qubit operations, control and coherence of P$_2^+$ charge
qubits.  Our results indicate that, although some of the operations may be
implemented as originally conceived, the spin and charge qubits based on
donors in silicon pose immense challenges in terms of precise nanostructure
fabrications because of the degenerate nature of the silicon conduction band.
Further studies of fabrication and innovative alternative approaches are
imperative in order to fully realize the potential of donor-based QC
architectures. 

\begin{acknowledgments}
This work was supported by Conselho Nacional de Desenvolvimento Cient\'\i fico e 
Tecnol\'ogico (CNPq), Instituto do Mil\^enio de Nanoci\^encias  and 
Funda\c c\~ao Carlos Chagas Filho de Amparo \` a Pesquisa do Estado do Rio de Janeiro
(FAPERJ) in Brazil, Advanced Research and Development Authority (ARDA) and 
Laboratory for Physical Sciences - National Security Agency (LPS-NSA) 
at the University of Maryland, and by The Army Research Office - Advanced Research and 
Development Authority (ARO-ARDA) at the University at Buffalo and the University of Maryland.
\end{acknowledgments}


\begin{thebibliography}{99}

\bibitem{Shor} P.W. Shor, ``Polynomial-time algorithms for prime
factorization and discrete logarithms on a quantum computer,'' in: S.
Goldwasser ed., {\it Proceedings of the 35th Annual  Symposium on the
Foundations of Computer Science} (IEEE Computer Society, Los Alamitos, 1994),
124-134.

\bibitem{nielsen00} M.A. Nielsen and I.L. Chuang, ``Quantum computation and
quantum information'' (Cambridge Univ. Press, Cambridge, U.K., 2000).

\bibitem{Reviews} 
D.P. DiVincenzo, ``Quantum Computation,'' Science {\bf 270} (1995) 255; 
A. Ekert and R. Jozsa, ``Quantum computation and Shor's factoring algorithm,''
Rev. Mod. Phys. {\bf 68} (1996) 733-753;  
A. Steane, ``Quantum computing,'' Rep. Prog. Phys. {\bf 61} (1998) 117-173; 
C.H. Bennett and D.P. DiVincenzo, ``Quantum information and computation,''
Nature {\bf 404} (2000) 247-255.

\bibitem{error} P.W. Shor, ``Scheme for reducing decoherence in
quantum computer memory,'' {\it Phys. Rev A} {\bf 52} (1995), R2493-R2496;
A.M. Steane, Error correcting codes in quantum theory, Phys. Rev. Lett. {\bf
77} (1996) 793-797.

\bibitem{barenco} A. Barenco, C.H. Bennett, R. Cleve, D.P.DiVincenzo,
N.Margolus, P. Shor, T. Sleator, J.A.Smolin, and H. Weinfurter, ``Elementary gates for 
quantum computation'', Phys. Rev. A {\bf 52} (1995) 3457-3467.

\bibitem{chuang01} L.M.K. Vandersypen , M. Steffen, G. Breyta, C.S. Yannoni CS, M.H. Sherwood,and I.L. Chuang, 
``Experimental realization of Shor's quantum factoring algorithm using nuclear
magnetic resonance'',  Nature {\bf 414} (2001) 883-887. 

\bibitem{igorrmp} I. \v{Z}uti\'{c}, J. Fabian, and S. Das Sarma,
``Spintronics: Fundamentals and applications'', Rev. Mod. Phys. {\bf 76}
(2004) 323-410.

\bibitem{SSReviews} S. Das Sarma, J. Fabian, I. \v{Z}uti\'{c}, ``Spin
electronics and spin computation,'' Solid State Commun. {\bf 119} (2001) 207-215;
X. Hu and S. Das Sarma, ``Overview of spin-based quantum dot quantum
computation'', Phys. Stat. Sol. (b) {\bf 238} (2003) 360-365;
X. Hu, ``Quantum Dot Quantum Computing'', arXiv:cond-mat/0411012; 
S. Das Sarma, R. de Sousa, X. Hu, B. Koiller, 
``Spin quantum computation in silicon nanostructures'',
Solid State Commun. {\bf 113} (2005) 737-746.
 
\bibitem{Exch} D. Loss and D. P. DiVincenzo, ``Quantum computation with
quantum dots,'' Phys. Rev. A {\bf 57} (1998) 120-126.  

\bibitem{Imam} A. Imamoglu, D.D. Awschalom, G. Burkard, D.P.
DiVincenzo, D. Loss, M. Sherwin, and A. Small, ``Quantum information
processing using quantum dot spins and cavity QED,'' Phys. Rev.
Lett. {\bf 83} (1999) 4204-4207.

\bibitem{Kane} B.E. Kane, ``A silicon-based nuclear spin quantum computer'' 
Nature {\bf 393} (1998) 133-137.

\bibitem{Privman} V. Privman, I.D. Vagner, and G. Kventsel, ``Quantum
computation in quantum-Hall systems,'' {\it Phys. Lett. A} {\bf 239} (1998)
141-146.

\bibitem{Vrijen} R. Vrijen, E. Yablonovitch, K. Wang, H.W. Jiang, A.
Balandin, V. Roychowdhury, Tal Mor, and D.P. DiVincenzo, 
``Electron-spin-resonance transistors for quantum computing in
silicon-germanium heterostructures,'' 
Phys. Rev. A {\bf 62} (2000) 012306.

\bibitem{voyles} P.M. Voyles, D.A. Muller, J.L. Grazui, P.H. Citrin, and
H.-J.L. Grossmann, ``Atomic-scale imaging of individual dopant atoms and
clusters in highly n-type bulk Si,''  Nature {\bf 416} (2002) 826-829. 



\bibitem{SCT} {\em Single Charge Tunneling}, ed. by H. Grabert and M.H.
Devoret (Plenum, New York, 1992).

\bibitem{Kohn} W. Kohn, ``Shallow impurity states in silicon and germanium,''
in {\it Solid State Physics}, vol. 5, 
F. Seitz and D. Turnbull Ed. New York: Academic Press, 1957,  pp. 257-320. 

\bibitem{KCHD} B. Koiller, R.B. Capaz, X. Hu and S. Das Sarma, ``Shallow donor
wavefunctions and donor-pair exchange in silicon: Ab initio theory and
floating-phase Heitler-London approach'',  Phys. Rev. B {\bf 70} (2004) 115207.

\bibitem{Klimeck} G. Klimeck, R.C. Bowen, T.B. Boykin, C.S. -Lazaro, T.A.
Cwik, and A.Stoica, ``Si tight-binding parameters from genetic algorithm 
fitting'', Superlattices Microstruct {\bf 27} (2000) 77-88. 

\bibitem{martins04} A.S. Martins, R. B. Capaz, and B. Koiller, ``Electric
field 
control and adiabatic evolution of shallow-donor impurities in silicon'', 
Phys. Rev. B {\bf 69} (2004) 085320.


\bibitem{martins02} A.S. Martins, J.G. Menchero, R.B. Capaz, and B. Koiller, 
``Atomistic description of shallow levels in semiconductors'', Phys. Rev. B
{\bf 65} (2002) 245205.

\bibitem{Skinner} A.J. Skinner, M.E. Davenport, and B.E. Kane, ``Hydrogenic
spin quantum computing in silicon: A digital approach'', Phys. Rev. Lett.
{\bf 90} (2003) 087901.

\bibitem{HD} X. Hu and S. Das Sarma, ``Hilbert-space structure of a
solid-state quantum computer: Two-electron states of a double-quantum-dot
artificial molecule,'' Phys. Rev. A {\bf 61} (2000) 062301.

\bibitem{Obrien} J.L. O'Brien, S.R. Schofield, M.Y. Simmons,  
R.G. Clark, A.S. Dzurak, N.J. Curson, B.E. Kane, N.S. McAlpine, 
M.E. Hawley, and G.W. Brown, ``Towards the fabrication of phosphorus qubits
for a silicon quantum computer,'' 
Phys. Rev. B {\bf 64} (2001) 161401.

\bibitem{encapsulation} S.R. Schofield, N.J. Curson, M.Y. Simmons, F.J.
Rue\ss, T. Hallam, L. Oberbeck, and R.G. Clark, ``Atomically precise
placement of single dopants in Si'',   
Phys. Rev. Lett. {\bf 91} (2003) 136104.

\bibitem{implant} T.M. Buehler, R.P. McKinnon, N.E. Lumpkin, R. Brenner, 
D.J. Reilly, L.D. Macks, A.R. Hamilton, A.S. Dzurak, and R.G. Clark, 
``A self-aligned fabrication process for silicon quantum computer devices,'' 
Nanotechnology {\bf 13} (2002) 686-690.

\bibitem{schenkel03} T. Schenkel,  A. Persaud, S.J. Park, J. Nilsson, J.
Bokor, J.A. Liddle, R. Keller, D.H. Schneider, D.W. Cheng, and
D.E. Humphries, ``Solid state quantum computer development in silicon with
single ion implantation,'' Journal of Applied Physics {\bf 94} (2003) 7017-7024.

\bibitem{Andres} K. Andres, R.N. Bhatt, P. Goalwin, T.M. Rice, and R.E.
Walstedt, ``Low-temperature magnetic-susceptibility of Si-P in the
non-metallic region,'' Phys. Rev. B {\bf 24} (1981) 244-260.

\bibitem{KHD1} B. Koiller, X. Hu and S. Das Sarma, 
``Exchange in silicon-based quantum computer architecture,'' Phys. Rev. Lett.
{\bf 88} (2002) 027903. 

\bibitem{KHD2} B. Koiller, X. Hu and S. Das Sarma, ``Strain effects on silicon
donor exchange: Quantum computer architecture considerations,''  Phys. Rev. B
{\bf 66} (2002) 115201.

\bibitem{wellard03} C.J. Wellard, L.C.L. Hollenberg, F. Parisoli, L. Kettle, 
H.-S. Goan, J.A.L. McIntosh, and D.N. Jamieson, ``Electron exchange coupling
for single-donor solid-state spin qubits,'' Phys. Rev. B {\bf 68} (2003)
195209. 

\bibitem{slater} J. C. Slater, {\it Quantum Theory of Molecules and Solids},
vol. 1, McGraw-Hill, New York, 1963. 

\bibitem{KHDD} B. Koiller, X. Hu, H.D. Drew, and S. Das Sarma, ``Disentangling
the Exchange Coupling of Entangled Donors in the Silicon Quantum Computer
Architecture'', Phys. Rev. Lett. {\bf 90} (2003) 067401.

\bibitem{KH} B. Koiller and X. Hu, ``Nanofabrication aspects of silicon-based
spin quantum gates'', 
IEEE Transactions in Nanotechnology {\bf 4} (2004) 113-115.

\bibitem{Hu04} X. Hu, B. Koiller, and S. Das Sarma, ``Charge qubits in
semiconductor quantum computer architectures: Tunnel coupling and
decoherence'', arXiv:cond-mat/0412340 (2004), to appear in Phys. Rev. B.

\bibitem{Cole00} B.E. Cole, J.B. Williams, B.T. King, M.S. Sherwin, and
C. R. Stanley, ``Coherent manipulation of semiconductor quantum bits with 
terahertz radiation'', Nature {\bf 410} (2000) 60-63.

\bibitem{Hayashi} T. Hayashi, T. Fujusawa, H. D. Cheong, Y. H. Jeong, and Y.
Hirayama, ``Coherent Manipulation of Electronic States in a Double Quantum
Dot'', Phys. Rev. Lett. {\bf 91} (2003) 226804.

\bibitem{hollenberg1} L.C.L. Hollenberg, A. S. Dzurak, C. J. Wellard, A.R.
Hamilton, D.J. Reilly, G.J. Milburn, and R.G. Clark, 
``Charge-based quantum computing using single donors in semiconductors'', 
Phys. Rev. B {\bf 69} (2004) 113301.

\bibitem{hollenberg2} L.C.L. Hollenberg, C. J. Wellard, C.I. Pakes, and A.G.
Fowler, ``Single-spin readout for buried dopant semiconductor qubits'', 
Phys. Rev. B {\bf 69} (2004) 233301.

\bibitem{Ekert}A.K. Ekert and R. Jozsa, 
``Quantum computation and Shor's factoring algorithm'', 
Rev. Mod. Phys. {\bf 68} (1996) 733-753.

\bibitem{Tanamoto} T. Tanamoto, ``One- and two-dimensional N-qubit systems in
capacitively coupled quantum dots'', Phys. Rev. A {\bf 64} (2001) 062306;
``Quantum gates by coupled asymmetric quantum dots and controlled-NOT-gate
operation'', {\it ibid.} {\bf 61} (2000) 022305.

\bibitem{Barrett} S. D. Barrett and G. J. Milburn, 
``Measuring the decoherence rate in a semiconductor charge qubit'', 
Phys. Rev. B {\bf 68} (2003) 155307.

\bibitem{Fedichkin} L. Fedichkin and A. Fedorov, 
``Error rate of a charge qubit coupled to an acoustic phonon reservoir'', 
Phys. Rev. A {\bf 69} (2004) 032311.


\end{thebibliography}
\end{document}